\documentclass[runningheads]{llncs}
\usepackage{hyperref}
\usepackage{amsmath} 
\usepackage{graphicx}
\usepackage{amssymb}
\usepackage{dsserif}
\usepackage[marginal]{footmisc}

\usepackage{bbding}
\usepackage[figuresright]{rotating}
\usepackage[normalem]{ulem}
\useunder{\uline}{\ul}{}

\begin{document}

\title{Progressive Frequency-Aware Network for Laparoscopic Image Desmoking}
%
\author{Jiale Zhang \inst{} and
Wenfeng Huang\inst{} \and
Xiangyun Liao\inst{}\inst{(}\Envelope\inst{)} \and Qiong Wang\inst{}\inst{(}\Envelope\inst{)}}
\authorrunning{Zhang et al.}
\institute{Guangdong Provincial Key Laboratory of Computer Vision and Virtual Reality Technology, Shenzhen Institute of Advanced Technology, Chinese Academy of Sciences \newline \email{ \{xy.liao,wangqiong\}@siat.ac.cn}}
\maketitle              
\footnote{Jiale Zhang and Wenfeng Huang contribute equally to this work.\\}
\footnote{This research was supported by multiple grants, including: The National Key Research and Development Program of China (2020YFB1313900), National Natural Science Foundation of China (62072452), Shenzhen Science and Technology Program (JCYJ20200109115627045, JCYJ20220818101408019, JCYJ20200109115201707) and Regional Joint Fund of Guangdong (2021B1515120011).}
\begin{abstract}
Laparoscopic surgery offers minimally invasive procedures with better patient outcomes, but smoke presence challenges visibility and safety. Existing learning-based methods demand large datasets and high computational resources.
We propose the Progressive Frequency-Aware Network (PFAN), a lightweight GAN framework for laparoscopic image desmoking, combining the strengths of CNN and Transformer for progressive information extraction in the frequency domain.
PFAN features CNN-based Multi-scale Bottleneck-Inverting (MBI) Blocks for capturing local high-frequency information and Locally-Enhanced Axial Attention Transformers (LAT) for efficiently handling global low-frequency information. PFAN efficiently desmokes laparoscopic images even with limited training data. Our method outperforms state-of-the-art approaches in PSNR, SSIM, CIEDE2000, and visual quality on the Cholec80 dataset and retains only 629K parameters. Our code and models are made publicly available at: \href{https://github.com/jlzcode/PFAN}{https://github.com/jlzcode/PFAN}.

\end{abstract}
\section{Introduction}

Laparoscopic surgery provides benefits such as smaller incisions, reduced post-operative pain, and lower infection rates~\cite{Jaschinski2018}. The laparoscope, equipped with a miniature camera and light source, allows visualization of surgical activities on a monitor. However, visibility can be hindered by smoke from laser ablation and cauterization. Reduced visibility negatively impacts diagnoses, decision-making, and patient health during intraoperative imaging and image-guided surgery, and hampers computer vision algorithms in laparoscopic tasks such as depth estimation, surgical reconstruction and lesion identification.
Although smoke evacuation equipment is commonly used, its high cost and impracticality make image processing-based approaches a more attractive alternative~\cite{Tchaka2017}. However, traditional image processing algorithms have limitations in efficacy and can cause visual distortions. Approaches based on atmospheric scattering models inaccurately treat smoke as a homogeneous scattering medium, potentially leading to tissue misidentification and surgical accidents. End-to-end deep learning approaches show promise, but acquiring large training datasets is difficult and time-consuming, especially for medical applications. Moreover, most deep learning-based models have large parameter counts, making them unsuitable for resource-constrained medical devices. Laparoscopic models must be adaptable to various smoke concentrations and brightness levels, applicable across different surgical environments, lightweight, and effective with limited datasets. 

In this study, we propose the Progressive Frequency-Aware Net (PFAN), an efficient, lightweight model built within the generative adversarial networks (GANs) framework for laparoscopic smoke removal. We address smoke removal by focusing on the image frequency domain, integrating high-frequency and low-frequency features to translate smoke-filled images into clear and no-artifacts smoke-free images. With only 629K parameters, PFAN demonstrates remarkable laparoscopic image desmoking results.
In summary, the contributions of this work include:

(1) Our proposed PFAN model effectively combines CNN and ViT to take into account frequency domain features of laparoscopic images. PFAN employs the MBI (CNN-based) and LAT (ViT-based) components to sequentially extract high and low-frequency features from the images. This approach establishes a robust feature extraction framework by leveraging the CNNs' local high-frequency feature extraction capabilities and the Transformers' global low-frequency feature extraction strengths.

(2) Our work introduces two innovations to the PFAN model: the Multi-scale Bottleneck-Inverting (MBI) Block, which extracts local high-frequency features using a multi-scale inverted bottleneck structure, and the Locally-Enhanced Axial Attention Transformer (LAT), which efficiently processes global low-frequency information with Squeeze-Enhanced Axial Attention and Locally-Enhanced Feed Forward Layer.

(3) The lightweight PFAN model effectively removes smoke from laparoscopic images with a favorable performance-to-complexity balance. It is suitable for resource-constrained devices. Evaluation results indicate its superiority over state-of-the-art methods, highlighting its effectiveness in removing surgical smoke from laparoscopic images.
\begin{figure}[htp]
    \centering
    \vspace{-2em}
    \includegraphics[width=12cm]{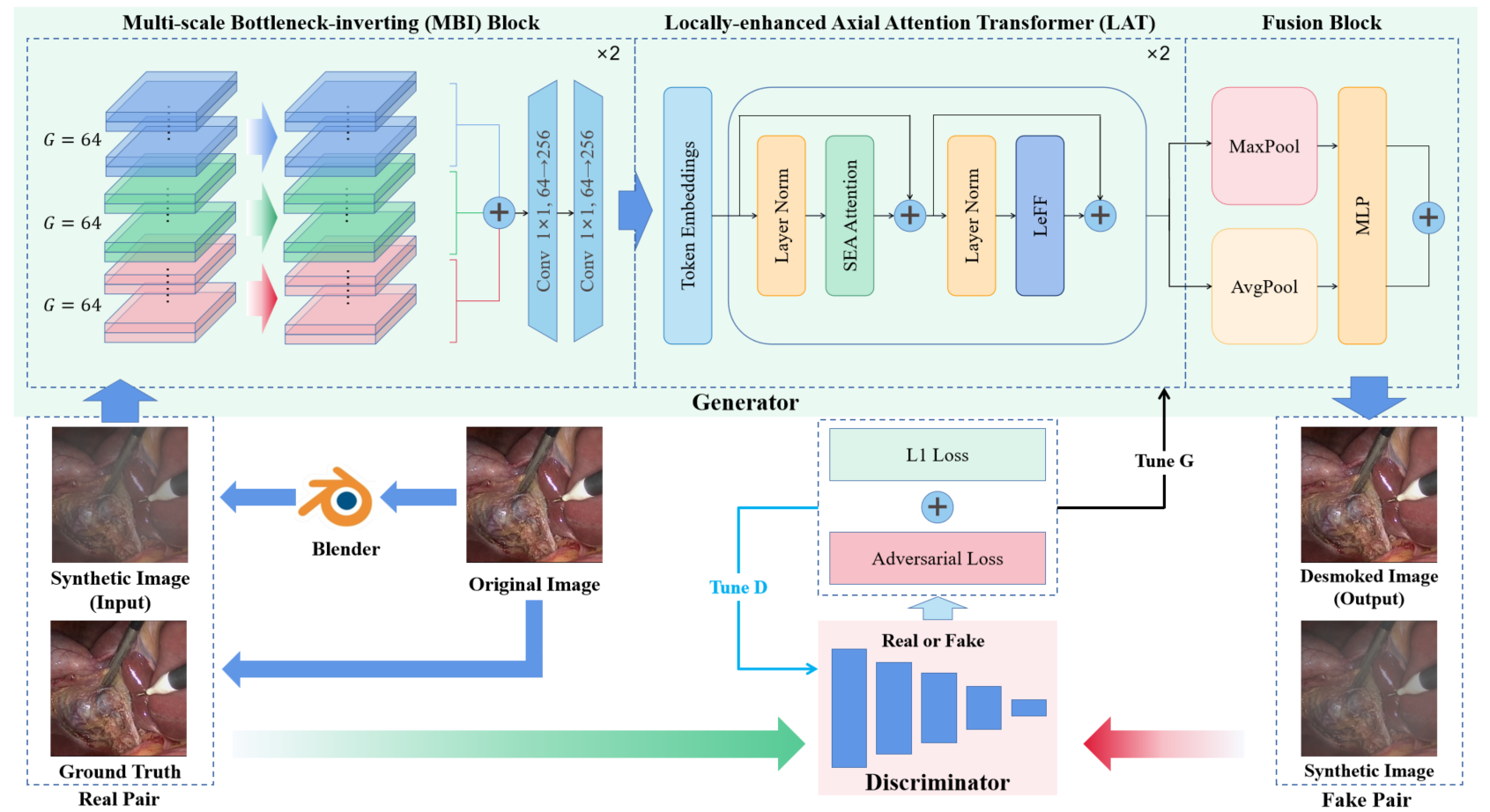}
    \caption{The flowchart of PFAN illustrates a framework consisting of a generator network (G) and a discriminator network (D). Within this proposed approach, the generator G incorporates Multi-scale Bottleneck-Inverting (MBI) Blocks and Locally-Enhanced Axial Attention Transformer (LAT) Blocks.}
    \label{fig:main}
    \vspace{-2em}
\end{figure}
\section{Related Work}
\subsection{Traditional Theory-Based Desmoking Methods}
Traditional desmoking techniques include image restoration and enhancement. Restoration methods, like the dark channel prior (DCP) by He et al.~\cite{He2009}, use atmospheric degradation and depth information, but face limitations in laparoscopic imaging. Enhancement techniques, such as Retinex algorithm~\cite{Rahman}, wavelet-based algorithms~\cite{Rong2014}, improving local contrast, increasing visibility and interpretability~\cite{Li2018}. Wang et al.~\cite{Wang2018a} created a variational desmoking approach, but it relies on assumptions regarding smoke's heterogeneous nature and varying depths.
\subsection{Deep Learning-Based Desmoking Methods}
Deep learning advances have fostered diverse frameworks for laparoscopic image smoke removal. Sabri et al.~\cite{Bolkar2018a} employed synthetic surgical smoke images with different smoke densities and utilized CNNs to remove smoke in a supervised setting, while DehazeNet~\cite{Removal2016} , AOD-Net~\cite{Li2017} and DM${^{2}}$F-Net~\cite{deng2019deep} relied on atmospheric scattering models, inappropriate for surgical environments. GANs~\cite{Goodfellow2020}, using game-theoretic approaches, generate realistic images. Techniques like Pix2Pix~\cite{Isola} employ conditional GANs for domain mapping. In medical imaging, GANs have been effective in PET-CT translation and PET image denoising~\cite{Armanious2016}. However, methods based on convolutional neural networks struggle with low-frequency information extraction, such as contour and structure. Vision Transformers (ViT)~\cite{Dosovitskiy2020} excel in low-frequency extraction, but their complexity restricts use in resource-limited medical devices. 

\section{Methodology}
 Fig.~\ref{fig:main} depicts our proposed PFAN, a lightweight CNN-ViT-based approach within a GAN architecture for desmoking in laparoscopic images, it extracts information progressively in the frequency domain by leveraging the strengths of CNNs and ViTs.  
In order to obtain the necessary corresponding smoky and non-smoky images, we integrate a graphics rendering engine into our learning framework to generate paired training data without manual labeling.
\subsection{Synthetic Smoke Generation}
\label{section:A}
We employ the Blender engine to generate smoke image pairs for model training, offering two advantages over physically-based haze formation models~\cite{Removal2016} and Perlin noise functions~\cite{Bolkar2018}. First, laparoscopic surgical smoke is localized and depth-independent, making traditional haze models unsuitable. Second, modern rendering engines provide realistic and diverse smoke shapes and densities using well-established, physically-based built-in models. With Blender's render engine, denoted by $\phi$, we generate the smoke evolution image sequence, $S_{Smoke}$, by adjusting parameters such as smoke source density, intensity, temperature, location ($S_{d}, S_{i}, S_{t}, S_{l}$), and light location and intensity ($L_{l}, L_{i}$):
\begin{equation}
S_{Smoke} = \phi (S_{d}, S_{i}, S_{t}, S_{l}, L_{l}, L_{i})
\end{equation}
Let $I_{Smoke}$ represent one frame of the smoke image sequence. To create a synthetic smoke evolution image sequence ($I_{Syn}$) within the surgical scene, we overlay a randomly generated frame of smoke evolution image sequence ($I_{Smoke}$) onto each smoke-free laparoscopic image ($I_{Smoke-free}$). The following formula represents this process: \begin{equation}
I_{Syn} = I_{Smoke-free} + I_{Smoke}
\end{equation}
The synthesized laparoscopic image sequence shows the evolution process of smoke. In the first frame of the synthesized image sequence, smoke is only present at a specific location within the image, simulating the situation of burning lesion areas in laparoscopic surgery. As time progresses, it disperses from the burning point outwards according to random density, temperature, and intensity parameters. The synthesis of an extensive range of realistic images depicting simulated surgical smoke is made possible through the utilization of a robust rendering engine. By incorporating variations in smoke, such as location, intensity, density, and luminosity, over-fitting is prevented in the network's training. 
\subsection{Multi-scale Bottleneck-Inverting (MBI) Block}
The MBI Block is designed to efficiently extract high-frequency features, drawing inspiration from various well-established neural networks~\cite{inception,Liu2022,Huang2022,huangijcnn}. Here, we denote input smoke images as ${ \mathcal{X}}_{Smoke} \in { \mathbb {R}}^{H\times W\times 3} $, and the set of high-frequency information extracted by each MBI Block can be defined as $  \lbrace{ { \mathcal{X} }_{HF} = {{ \mathcal{X} }_{{HF}_{1}}
, . . . , { \mathcal{X} }_{{HF}_{k}}}
}\rbrace$ . Within each MBI Block, group convolution is represented as GConv, and the multi-scale feature can be obtained as: 
\begin{equation}
{ \mathcal{X} }_{MS} = GConv_{i,g}({ \mathcal{X} }_{Smoke}) + GConv_{j,g}({ \mathcal{X} }_{Smoke}) + GConv_{k,g}({ \mathcal{X} }_{Smoke}) 
\end{equation}
Here, $i$, $j$, and $k$ represent the size of the receptive field, which were set to 3, 7, and 11, respectively. We choose GELU~\cite{Hendrycks2016} instead of RELU as the activation function following each convolution layer, given its smoother properties and proven higher performance. 
The parameter $g$ indicates that, during group convolution, input features will be divided into $g$ groups. In this paper, this value is set to 64, which matches the feature channels, resulting in a significant reduction of parameters by 1/64 in comparison to standard convolution. Next, we merge the multi-scale feature ${ \mathcal{X} }_{MS}$ and expand it to a high-dimensional representation using point-wise convolution. Following this, features are projected back to a low-dimensional representation through point-wise convolution, represented as 
\begin{equation}
{ \mathcal{X} }_{HF} = PwConv_{high\rightarrow low}(PwConv_{low\rightarrow high}({ \mathcal{X} }_{MS}))
\end{equation}
\subsection{Locally-Enhanced Axial Attention Transformer (LAT) Block}
Applying ViT models to desmoke laparoscopic images faces challenges. ViT's multi-head self-attention layer applies global attention, neglecting differing frequencies and local high-frequency information. Additionally, ViT's computational cost increases quadratically with token count, limiting its use with high-resolution feature maps. To overcome these issues, we introduce the Locally-Enhanced Axial Attention Transformer (LAT) Block. It combines streamlined squeeze Axial attention for global low-frequency semantics and a convolution-based enhancement branch for local high-frequency information. The LAT Block captures long-range dependencies and global low-frequency information with low parameter counts.

Given the features at MBI Block outputs, ${ \mathcal{X}}_{MBI}$, LAT first reshapes the input into patch sequences using ${H\times H}$ non-overlapping windows. And then Squeeze-Enhanced Axial Attention computes attention maps (${ \mathcal{X} }_{Sea}$) for each local window. To further process the information, LAT replaces the multi-layer perceptron (MLP) layers in a typical ViT with a Locally-Enhanced Feed Forward Layer. Additional skip connections enable residual learning and produce ${ \mathcal{X} }_{LAT}$. 
\begin{equation}
{ \mathcal{X} }_{Sea} = SEA(LN({\mathcal{X}}_{MBI}))+{\mathcal{X}}_{MBI},
\quad { \mathcal{X} }_{LAT} = LEFF(LN({\mathcal{X}}_{Sea}))+{\mathcal{X}}_{Sea}
\end{equation}
Here, $ \mathcal{X}_{Sea}$ and $ \mathcal{X}_{LAT}$ correspond to the outputs of the Squeeze-Enhanced Axial Attention and LEFF modules, respectively. LN denotes layer normalization~\cite{Kotwal2016}. We discuss Squeeze-Enhanced Axial Attention and LEFF in detail in subsequent sections.
\begin{figure}[htp]
    \centering
                \vspace{-2em}
    \includegraphics[width=12cm]{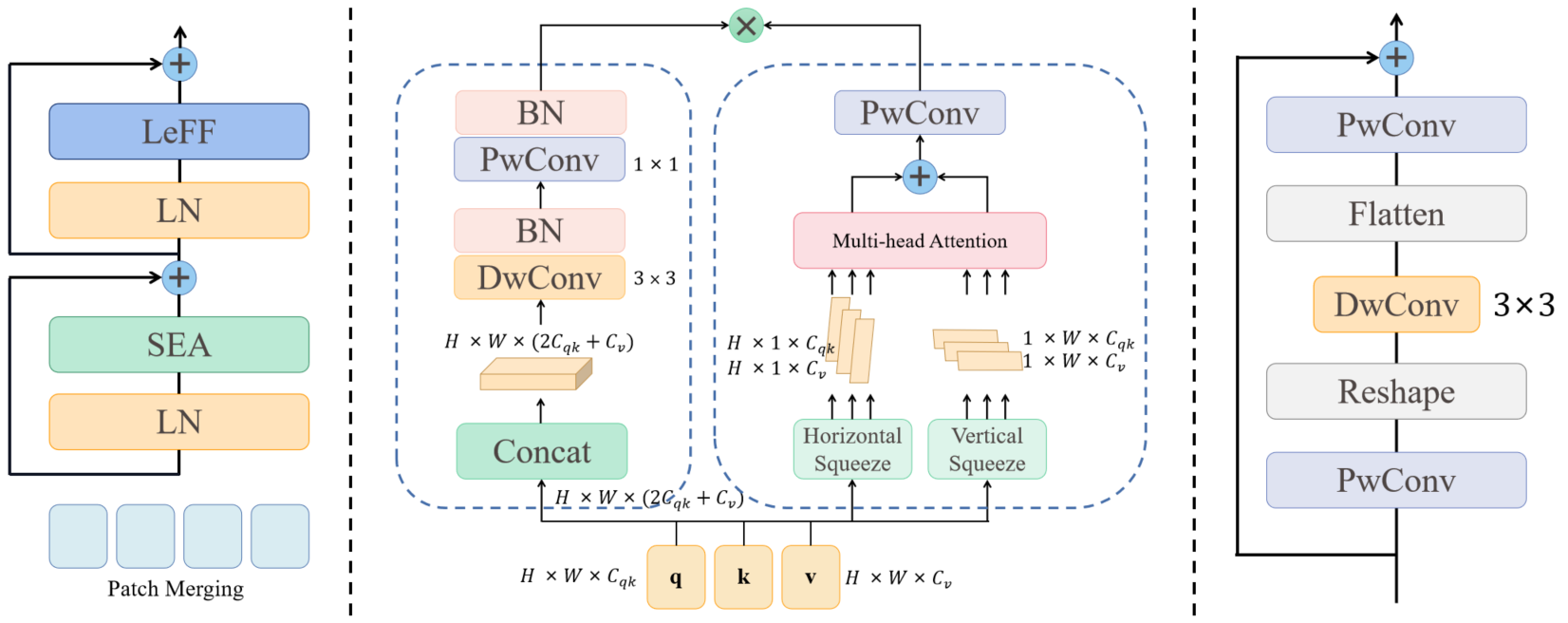}
    \caption{Left: the schematic illustration of the proposed Locally-Enhanced Axial Attention Transformer Block. Middle: Squeeze-Enhanced Axial Attention Layer. Right:  Locally-Enhanced feed-forward network. }
    \label{fig:lat}
    \vspace{-2em}
\end{figure}
\subsubsection{Squeeze-Enhanced Axial Attention (SEA)} 
The Squeeze-Enhanced Axial Attention utilized in the Locally-Enhanced Axial Attention Transformer (LAT) is designed to extract global information in a succinct way.
Initially, we compute ${\mathbf q}$, ${\mathbf k}$, and ${\mathbf v}$ by 
$
{\mathbf q} = W_{q} *{ \mathcal{X} } ,
{\mathbf k} = W_{k} *{ \mathcal{X} } ,
{\mathbf v} = W_{v} *{ \mathcal{X} }  ,
$
where ${ \mathcal{X} }  \in \mathbb {R}^{H \times W \times C}$. $W_{q},W_{k} \in \mathbb {R}^{C_{qk} \times C}$ and $W_{v} \in \mathbb {R}^{C_{v} \times C}$ are learnable weights.
Then, a horizontal squeeze $ {\mathbf q}_{h} $ is executed by averaging the query feature map along the horizontal direction and a vertical squeeze $ {\mathbf q}_{v}$ is applied in the vertical direction. 
\begin{equation}
\begin{split}
{\mathbf q}_{h} = \frac{1}{W}\left({\mathbf q}^{(C_{qk},H,W)}  {\mathbb{1}}_{W} \right)^{\rightarrow(H,C_{qk})} ,
\quad{\mathbf q}_{h} = \frac{1}{H}\left({\mathbf q}^{(C_{qk},W,H)}  {\mathbb{1}}_{H} \right)^{\rightarrow(W,C_{qk})} 
\end{split}
\end{equation}
The notation ${\mathbf z}^{\rightarrow(\cdot)}$ represents the permutation of tensor ${\mathbf z}$'s dimensions, and $ {\mathbb{1}}_{m} \in {\mathbb{R}}_{m} $ is a vector with all elements equal to 1. The squeeze operation on ${\mathbf q}$ is also applied to ${\mathbf k}$ and ${\mathbf v}$, resulting in ${\mathbf q}_{h},{\mathbf k}_{h},{\mathbf v}_{h} \in {\mathbb{R}}^{H\times C_{qk}}, {\mathbf q}_{v},{\mathbf k}_{v},{\mathbf v}_{v} \in {\mathbb{R}}^{W\times C_{qk}}$. The squeeze operation consolidates global information along a single axis, thereby significantly improving the subsequent global semantic extraction process, as demonstrated by the following equation.
\begin{equation}
\label{seafun}
\begin{split}
{\mathbf y}_{(i,j)} = \sum_{p=1}^H softmax_{p} \left( \mathbf{q}_{h}^{{i}{\mathsf{T}} }\mathbf{k}_{h}^{p}\right)\mathbf{v}_{h}^{p} + \sum_{p=1}^W softmax_{p} \left(  \mathbf{q}_{v}^{{j}{\mathsf{T}}} \mathbf{k}_{v}^{p} \right) \mathbf{v}_{v}^{p}
\end{split}
\end{equation}
As can be seen, in Squeeze-Enhanced Axial Attention, each position of the feature map only propagates information in two squeezed axial features, while in traditional self-attention (as in the following equation), each position of the feature map calculates self-attention with all positions.
\begin{equation}
\label{safun}
\begin{split}
{\mathbf y}_{(i,j)} = \sum_{p \in \mathcal{G}_{(i,j)} }softmax_{p} \left( \mathbf{q}_{(i,j)}^{\mathsf{T}} \mathbf{k}_{p}\right)\mathbf{v}_{p}
\end{split}
\end{equation}
The traditional global self-attention is as above, where $ \mathcal{G}(i,j) $ means all positions on the feature map at location $(i, j)$.  When a conventional attention module is applied to a feature map with dimensions $ H\times W\times C $ the time complexity becomes $ O(H^{2}W^{2}(C_{qk}+C_{v}))$, resulting in low efficiency.
However, with SEA, the time complexity for squeezing $q$, $k$, $v$ is $ O((H+W)(2C_{qk}+C_{v}))$ and the attention operation takes $ O((H^{2}+W^{2})(C_{qk}+C_{v}))$ 
 time. Consequently, our squeeze Axial attention successfully lowers the time complexity to $ O(HW)$, ensuring a more efficient and faster process.

\subsubsection{Locally-Enhanced Feed-Forward Network (LEFF)}
Adjacent pixels play a crucial role in image desmoking, as demonstrated in~\cite{Wang2018a}, which highlights their essential contribution to image dehazing and denoising.
However, previous research~\cite{Tchaka2017} has highlighted the limited ability of the Feed-Forward Network (FFN) within the standard Transformer to effectively utilize local context. To address this limitation, we introduce a depth-wise convolutional block to LAT, inspired by recent studies~\cite{Li2021c}. As depicted in Fig.~\ref{fig:lat} (Right), we begin by applying a linear projection layer to each token to augment its feature dimension. Subsequently, we reshape the tokens into 2D feature maps and implement a ${3 \times 3}$ depth-wise convolution to capture local information. Afterward, we flatten the features back into tokens and reduce the channels using another linear layer to align with the input channel dimensions. $LeakyReLU$ serves as the activation function following each linear or convolution layer.

\subsubsection{Fusion Block}
We employ Channel Attention~\cite{woo2018cbam} as the Fusion Block in our approach to enhance the cross-channel feature fusion capabilities. The Channel Attention mechanism models inter-dependencies between channels of features, enabling adaptive adjustment of feature responses across different channels, and assigning corresponding weights. Embedding channel attention can facilitate adaptive enhancement and fusion of convolution and corresponding Transformer features in the LAT module. The attention map, $ \mathcal{X}_{CA} $, can be calculated using the function, where $\sigma$ represents the $Sigmoid$ function.
\begin{equation}
\mathcal{X}_{CA} = \sigma \left(LEFF(AvgPool(\mathcal{X}_{LAT})) + LEFF(MaxPool(\mathcal{X}_{LAT}))\right) 
\end{equation}
Afterward, the low-frequency information $\mathcal{X}_{LF}$ is acquired as described in (\ref{xlf}). 
\begin{equation}
\label{xlf}
\begin{split}
\mathcal{X}_{LF} = \mathcal{X}_{LAT} \cdot \mathcal{X}_{CA}
\end{split}
\end{equation}
To achieve the smoke-free result,  $\mathcal{X}_{Smoke-free}$, the low-frequency information of the original input smoke image $\mathcal{X}_{LF}$ is combined with the high-frequency information $\mathcal{X}_{HF}$, which is the output of MBI blocks.
\begin{equation}
\begin{split}
\mathcal{X}_{Smoke-free} = \mathcal{X}_{HF} + \mathcal{X}_{LF}
\end{split}
\end{equation}
\section{Experiment}
\subsection{Data Collections}
We used images from the Cholec80 dataset~\cite{Twinanda2017}, consisting of 80 cholecystectomy surgery videos by 13 surgeons.  We sampled 1,500 images at 20-second intervals from these videos, selecting 660 representative smoke-free images. As detailed in Section~\ref{section:A}, we added synthetic random smoke, yielding 660 image pairs, divided in an 8:1:2 ratio for training,  validation, and testing. Synthetic smoky images were generated according to Section~\ref{section:A}. Importantly, each dataset contained distinct videos, ensuring no overlap.

\subsection{Implementation Details}

Our experiments utilized six NVIDIA RTX 2080Ti GPUs. Initially, we trained the Discriminator (PatchGAN) for one epoch to provide a rough smoke mask, followed by iterative training of the Discriminator and Generator while freezing the PatchGAN's parameters during the Generator's training. We employed an Adam solver with a learning rate of 0.0002, momentum parameters $\beta1$ = 0.5 and $\beta2$ = 0.999, and a batch size of 6. Consistent with prior research, random cropping was used for generating training and validation patches.

\begin{figure}[h]
    \centering
    \vspace{-2em}
    \includegraphics[width=1\linewidth]{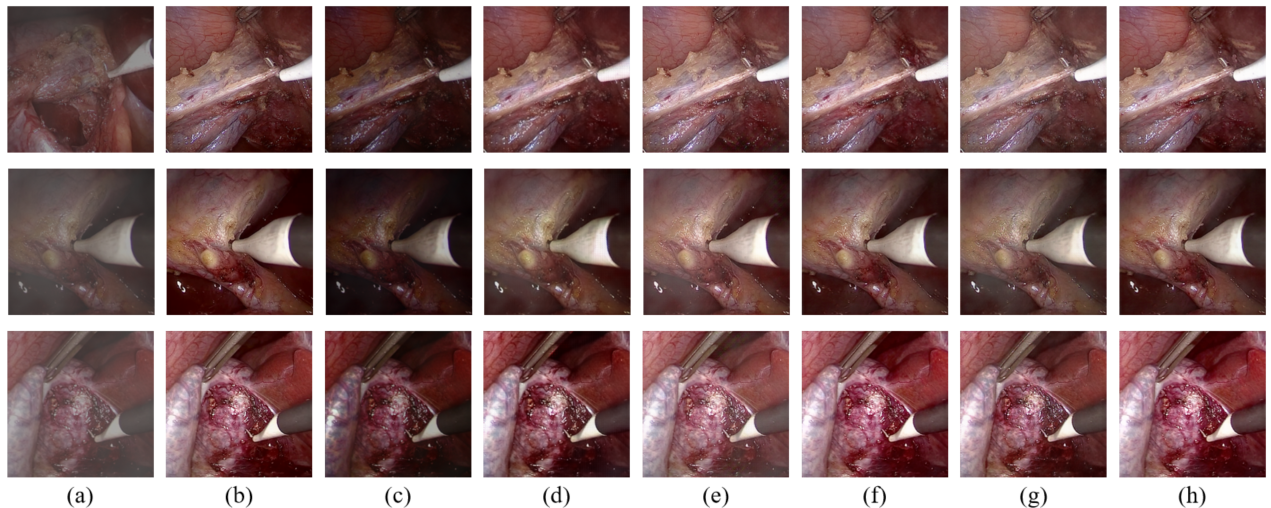}
    \caption{Comparison experiments between SOTAs. (a) Input  (b) Ground Truth, (c) Dark Channel Prior(DCP)~\cite{He2009} (d)  CycleGAN + ResNet, (e) CycleGAN + U-Net, (f)  Pix2Pix + ResNet, (g) Pix2Pix + U-Net, and (h) Ours.}
    \label{fig:compare}
    \vspace{-2em}
\end{figure}

\section{Result}

\begin{table*}[htb]
\centering
\vspace{-2em}
\caption{Quantitative results. The best and second-best results are highlighted and
underlined, respective}
\label{CMP}
\begin{tabular}{|llcccr|}
\hline
\multicolumn{2}{|c|}{Model}                                     & \multicolumn{1}{c|}{Parameters$\downarrow$} & \multicolumn{1}{c|}{PSNR$\uparrow$}             & \multicolumn{1}{c|}{SSIM$\uparrow$}            & CIEDE2000$\downarrow$       \\ \hline \hline
\multicolumn{2}{|c|}{DCP}                                       & \multicolumn{1}{c|}{/}          & \multicolumn{1}{c|}{27.6250}          & \multicolumn{1}{c|}{0.5528}          & 35.9952         \\ \hline
\multicolumn{1}{|l|}{CycleGAN} & \multicolumn{1}{l|}{U-Net} & \multicolumn{1}{c|}{54414K}     & \multicolumn{1}{c|}{28.7449}          & \multicolumn{1}{c|}{0.7621}          & 10.3298         \\ \hline
\multicolumn{1}{|l|}{CycleGAN} & \multicolumn{1}{l|}{ResNet6}   & \multicolumn{1}{c|}{7841K}      & \multicolumn{1}{c|}{29.0250}          & \multicolumn{1}{c|}{0.7826}          & 9.5821          \\ \hline
\multicolumn{1}{|l|}{CycleGAN} & \multicolumn{1}{l|}{ResNet9}   & \multicolumn{1}{c|}{11383K}     & \multicolumn{1}{c|}{29.0926} & \multicolumn{1}{c|}{0.7802} & 9.2868          \\ \hline
\multicolumn{1}{|l|}{Pix2Pix}  & \multicolumn{1}{l|}{U-Net} & \multicolumn{1}{c|}{54414K}     & \multicolumn{1}{c|}{29.2967}          & \multicolumn{1}{c|}{0.7073}          & 8.8060          \\ \hline
\multicolumn{1}{|l|}{Pix2Pix}  & \multicolumn{1}{l|}{ResNet6}   & \multicolumn{1}{c|}{7841K}      & \multicolumn{1}{c|}{29.8249}          & \multicolumn{1}{c|}{0.8358}          & 6.9364          \\ \hline
\multicolumn{1}{|l|}{Pix2Pix}  & \multicolumn{1}{l|}{ResNet9}   & \multicolumn{1}{c|}{11383K}     & \multicolumn{1}{c|}{29.8721}          & \multicolumn{1}{c|}{0.8417}          & {\ul 6.7046}          \\ \hline 
\multicolumn{1}{|l|}{Pix2Pix}  & \multicolumn{1}{l|}{Uformer}   & \multicolumn{1}{c|}{85605K}     & \multicolumn{1}{c|}{29.7030}          & \multicolumn{1}{c|}{0.8026}          & { 8.0602}          \\ \hline \hline
\multicolumn{6}{|c|}{Ablation Models} \\ \hline \hline
\multicolumn{2}{|l|}{w/o Multi-scale}                           & \multicolumn{1}{c|}{613K}       & \multicolumn{1}{c|}{\ul 29.9970}          & \multicolumn{1}{c|}{0.8692}          & 6.9362          \\ \hline
\multicolumn{2}{|l|}{w/o Fusion Block}                             & \multicolumn{1}{c|}{629K}       & \multicolumn{1}{c|}{29.4425}          & \multicolumn{1}{c|}{0.7814}          & 8.1200          \\ \hline
\multicolumn{2}{|l|}{w/o MBI}                                   & \multicolumn{1}{c|}{540K}       & \multicolumn{1}{c|}{29.7599}          & \multicolumn{1}{c|}{\ul 0.9029}          & 6.9149          \\ \hline
\multicolumn{2}{|l|}{w/o LAT}                                   & \multicolumn{1}{c|}{90K}        & \multicolumn{1}{c|}{28.8936}          & \multicolumn{1}{c|}{0.7857}          & 10.1284         \\ \hline \hline
\multicolumn{2}{|l|}{Ours}                                      & \multicolumn{1}{c|}{629K}       & \multicolumn{1}{c|}{\textbf{30.4873}} & \multicolumn{1}{c|}{\textbf{0.9061}} & \textbf{5.4988} \\ \hline
\end{tabular}
\vspace{-2em}
\end{table*}

In our quantitative evaluations, we assess desmoking performance by comparing smoke-free images to their desmoked counterparts using the following metrics: number of Parameters, Peak Signal-to-Noise Ratio (PSNR), Structural Similarity Index (SSIM)~\cite{Hore2010}, and CIEDE2000~\cite{luo2001development} (which represents color reconstruction accuracy for the human visual system). 
We compare the proposed method with eight state-of-the-art desmoking and dehazing methods including both a traditional image processing approach (DCP \cite{He2009}) and the most recent deep learning-based methods
(original CycleGAN~\cite{Liu2020} (with U-Net~\cite{ronneberger2015u}), CycleGAN with ResNet~\cite{he2016deep} (6Blocks), CycleGAN with ResNet (9Blocks), original Pix2Pix~\cite{Isola} (with U-Net), Pix2Pix with ResNet (6Blocks), Pix2Pix with ResNet (9Blocks), Pix2Pix with Uformer~\cite{Uformer}).

Table~\ref{CMP} demonstrates our model's superior performance compared to alternative methods based on synthetic datasets. The highest PSNR and SSIM values, and the lowest CIEDE2000 value, emphasize our approach's effectiveness in smoke removal tasks. Fig.~\ref{fig:compare} presents a subjective evaluation of desmoking results, emphasizing previous approaches' limitations in adequately removing smoke. Non-deep learning methods often produce low-brightness, color-shifted images due to DCP's unsuitability for surgical applications with complex lighting and varied smoke. Although deep learning techniques better restore brightness, CycleGAN and Pix2Pix cannot fully eliminate smoke, as evidenced by residual smoke in some image portions (Fig.~\ref{fig:compare}). These methods also result in unclear tissue contours due to CNN-based models' restricted global low-frequency feature extraction. In contrast, our methodology yields cleaner images with enhanced brightness, sharp details, and distinct edges.

\subsection{Evaluation under Different Smoke Densities}
\begin{figure}[htb]
    \centering
    \vspace{-2em}
    \includegraphics[width=1\linewidth]{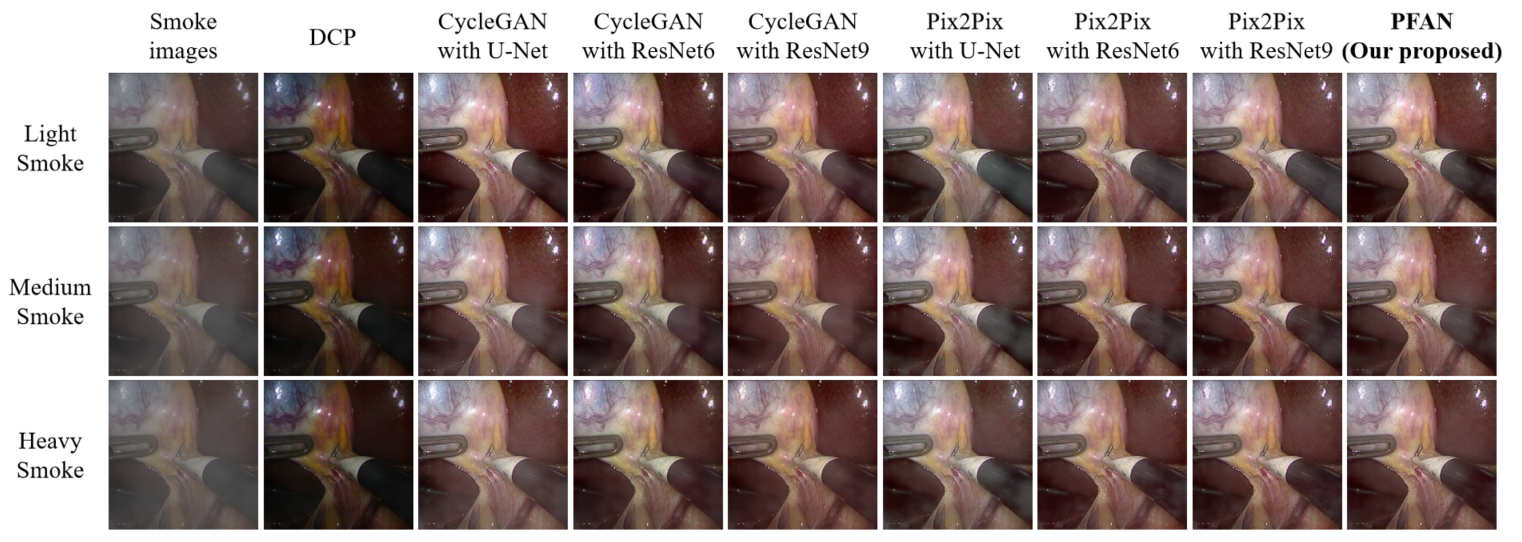}
    \caption{Qualitative comparison between SOTAs under different smoke densities}
    \label{fig:densitycmp}
    \vspace{-2em}
\end{figure}


Smoke impairs image information, often irreversibly, depending on thickness. To evaluate networks' desmoking performance at varying densities, we analyzed light, medium, and heavy smoke levels. We generated test sets for each density level with fixed starting positions and temperatures. Fig.\ref{fig:densitycmp} displays rendered smoke images ($I_{Syn}$) and desmoked results from five methods and our approach. DCP struggles to restore dark-red tissue colors, whereas deep learning-based techniques perform better using context. Pix2Pix produces similar results but falters for some images, introducing artificial reflections. Our method achieves clean results with minor saturation deviations, even under dense smoke conditions. Table \ref{Density} compares our approach to five alternatives, consistently yielding the highest SSIM and PSNR while reducing CIEDE2000, outperforming other established methods.

\begin{table*}[htb]
\caption{Quantitative comparison between SOTAs under different smoke densities}
\label{Density}
\resizebox{1\textwidth}{!}{
\begin{tabular}{|ll|ccc|ccc|ccr|}
\hline
\multicolumn{2}{|c|}{Smoke Density}       & \multicolumn{3}{c|}{Light Smoke}                                                                 & \multicolumn{3}{c|}{Medium Smoke}                                                              & \multicolumn{3}{c|}{Heavy Smoke}                                                               \\ \hline
\multicolumn{2}{|c|}{Model}               & \multicolumn{1}{c|}{PSNR$\uparrow$}             & \multicolumn{1}{c|}{SSIM$\uparrow$}            & CIEDE2000$\downarrow$       & \multicolumn{1}{c|}{PSNR$\uparrow$}             & \multicolumn{1}{c|}{SSIM$\uparrow$}            & CIEDE2000$\downarrow$       & \multicolumn{1}{c|}{PSNR$\uparrow$}             & \multicolumn{1}{c|}{SSIM$\uparrow$}            & CIEDE2000$\downarrow$       \\ \hline
\multicolumn{2}{|c|}{DCP}                 & \multicolumn{1}{c|}{27.6611}          & \multicolumn{1}{c|}{0.6215}          & 30.1270         & \multicolumn{1}{c|}{27.6811}          & \multicolumn{1}{c|}{0.5887}          & 32.9143         & \multicolumn{1}{c|}{27.6944}          & \multicolumn{1}{c|}{0.5807}          & 33.8072         \\ \hline
\multicolumn{1}{|l|}{CycleGAN} & U-Net & \multicolumn{1}{c|}{29.0426}          & \multicolumn{1}{c|}{0.7778}          & 8.5370          & \multicolumn{1}{c|}{28.9490}          & \multicolumn{1}{c|}{0.7607}          & 10.7167         & \multicolumn{1}{c|}{28.8837}          & \multicolumn{1}{c|}{0.7639}          & 10.7521         \\ \hline
\multicolumn{1}{|l|}{CycleGAN} & ResNet6  & \multicolumn{1}{c|}{29.0713}          & \multicolumn{1}{c|}{0.7958}          & 8.2868          & \multicolumn{1}{c|}{28.7621}          & \multicolumn{1}{c|}{0.7741}          & 11.7635         & \multicolumn{1}{c|}{28.7647}          & \multicolumn{1}{c|}{0.7755}          & 11.6661         \\ \hline
\multicolumn{1}{|l|}{CycleGAN} & ResNet9  & \multicolumn{1}{c|}{29.3232}          & \multicolumn{1}{c|}{0.8002}          & 7.8017          & \multicolumn{1}{c|}{28.7466}          & \multicolumn{1}{c|}{0.7650}          & 11.9671         & \multicolumn{1}{c|}{28.9379}          & \multicolumn{1}{c|}{0.7711}          & 10.8202         \\ \hline
\multicolumn{1}{|l|}{Pix2Pix}  & U-Net & \multicolumn{1}{c|}{29.2652}          & \multicolumn{1}{c|}{0.7270}          & 8.9004          & \multicolumn{1}{c|}{29.4071}          & \multicolumn{1}{c|}{0.7119}          & 9.1812          & \multicolumn{1}{c|}{29.4474}          & \multicolumn{1}{c|}{0.7199}          & 8.9037          \\ \hline
\multicolumn{1}{|l|}{Pix2Pix}  & ResNet6  & \multicolumn{1}{c|}{{\ul 29.9776}}    & \multicolumn{1}{c|}{0.8404}          & {\ul 6.6498}    & \multicolumn{1}{c|}{{\ul 30.1833}}    & \multicolumn{1}{c|}{{\ul 0.8288}}    & 6.8033          & \multicolumn{1}{c|}{30.2138}          & \multicolumn{1}{c|}{0.8344}          & {\ul 6.2970}    \\ \hline
\multicolumn{1}{|l|}{Pix2Pix}  & ResNet9  & \multicolumn{1}{c|}{29.9492}          & \multicolumn{1}{c|}{{\ul 0.8484}}    & 6.6610          & \multicolumn{1}{c|}{30.1498}          & \multicolumn{1}{c|}{0.8372}          & \textbf{6.7079} & \multicolumn{1}{c|}{{\ul 30.3287}}    & \multicolumn{1}{c|}{{\ul 0.8434}}    & 6.7079          \\ \hline
\multicolumn{2}{|c|}{Ours}                & \multicolumn{1}{c|}{\textbf{30.1209}} & \multicolumn{1}{c|}{\textbf{0.8856}} & \textbf{6.5182} & \multicolumn{1}{c|}{\textbf{30.2740}} & \multicolumn{1}{c|}{\textbf{0.8704}} & {\ul 6.8001}    & \multicolumn{1}{c|}{\textbf{30.5223}} & \multicolumn{1}{c|}{\textbf{0.8762}} & \textbf{6.1147} \\ \hline
\end{tabular}}
\vspace{-2em}
\end{table*}

\subsection{Ablation Studies}
\begin{figure}[b]
    \vspace{-2em}
    \centering
    \includegraphics[width=12cm]{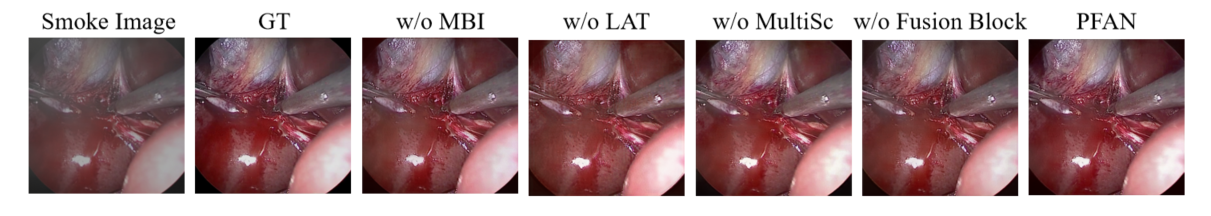}
    \caption{Qualitative results of ablation experiments.}
    \label{fig:ablationimg}
    \vspace{-2em}
\end{figure}
We design a series of ablation experiments to analyze the effectiveness of each of the modules we propose. The ablation results are reported in Table ~\ref{CMP}.\\
\textbf{Effectiveness of the MBI Block:}
The goal of the MBI Block is to effectively capture multi-scale, high-frequency details. Fig. \ref{fig:ablationimg} demonstrates that removing the MBI Block results in remaining smoke and blurry edges and textures in some image portions. This limitation in high-frequency detail extraction makes it challenging to obtain satisfactory desmoking outcomes. In Table. \ref{CMP}, our PFAN outperforms the model without the MBI Block in terms of PSNR, SSIM, and CIEDE2000 metrics. This comparison highlights the critical role of MBI Blocks in achieving superior results.\\
\textbf{Effectiveness of the LAT Block:}
The ViT-based LAT Blocks aim to extract global low-frequency information. Fig.~\ref{fig:ablationimg} shows that the model without LAT Blocks achieves a visually similar desmoking effect to the Ground Truth (GT); however, the color appears dull and exhibits noticeable distortion compared to the original smoke-free image. The higher CIEDE2000 value indicates insufficient low-frequency feature extraction. Furthermore, the lower PSNR and SSIM values demonstrate the effectiveness of the LAT module.\\
\textbf{Effectiveness of the Multi-scale MBI:}
Our approach employs group convolution with varying receptive fields in the MBI Block, facilitating multi-scale high-frequency information extraction. We conducted an ablation study, replacing multi-scale convolutions in the MBI Block with only $3\times 3$ group convolutions. Fig. \ref{fig:ablationimg} reveals substantial improvements in smoke removal, but the tissue in the central scalpel area appears blurred. Table. \ref{CMP} demonstrates the ``w/o Multi-scale" model achieves comparable performance to PFAN in terms of CIEDE2000 and PSNR; however, the SSIM value is significantly inferior, highlighting the importance of Multi-scale group convolutions in the MBI Block.\\
\textbf{Effectiveness of Fusion Block:}
The Fusion Block in our proposed method leverages channel attention for adaptive discriminative fusion between image Transformer features and convolutional features, enhancing the network's learning capability. Importantly, omitting channel attention leads to the most significant decline in SSIM value among the four ablation experiments. Additionally, noticeable differences in both PSNR and CIEDE2000 emerge compared to the PFAN results, underscoring channel attention's crucial role in PFAN.

\section{Limitations}
Our method has a few limitations. It overlooks external factors such as water vapor and pure white gauze, which can degrade image quality and then impede desmoking performance. Future iterations should incorporate these elements into training and testing to ensure clinical applicability. Moreover, our proposed single-frame 
 desmoking method may introduce temporal discontinuity in video desmoking tasks due to smoke density fluctuations. Thus, based on our current method, further investigation into spatial-temporal convolution techniques is necessary for enhancing laparoscopic video desmoking.

\section{Conclusion}

In conclusion, we proposed a groundbreaking deep learning method PFAN for laparoscopic image desmoking. By incorporating the lightweight and efficient CNN-ViT-based approach with the innovative CNN-based Multi-scale Bottleneck-Inverting (MBI) Blocks and Locally-Enhanced Axial Attention Transformers (LAT), PFAN effectively captures both low and high-frequency information for desmoking analysis, even with a limited dataset. The evaluation on the synthetic Cholec80 dataset, with various smoke-dense images, showcases the superiority of PFAN compared to existing SOTAs in performance and visual effects. Additionally, PFAN maintains a lightweight design, making it a feasible and desirable choice for implementation in medical equipment. 
Our desmoking method enables advanced applications. It enhances surgical safety by providing real-time desmoked images, serving as a valuable reference during ablation procedures. Furthermore, beyond aiding surgeons directly, the technology can also improve the robustness of various vision-based surgical assistance systems when used as a preprocessing step.

\bibliographystyle{splncs04}

\end{document}